# Comparing modern techniques for querying data starting from top-k and skyline queries

Fabio Patella

Politecnico di Milano
Milan, Italy
fabio.patella@polimi.it

**Abstract**

To make intelligent decisions over complex data by discovering a set of interesting options is something that has become very important for users of modern applications. Consequently, researchers are studying new techniques to overcome limitations of traditional ways of querying data from databases as top-k queries and skyline queries. Over the past few years new methods have been developed as Flexible Skylines, Regret Minimization and Skyline ordering/ranking. The aim of this survey is to describe these techniques and some their possible variants comparing them and explaining how they improve traditional methods.

## 1 Introduction

A problem that concerns needs of users in modern applications is the multi-objective optimization that is the simultaneous optimization of different criteria based on objects in a dataset in order to return to the user the best possible objects for her and her criteria. Two different traditional approaches for that are top-k-queries and skyline queries. The idea of top-k queries is to rank and then return the best k tuples of a dataset, in which each tuple has a certain fixed number of attributes, through an objective function of attributes of tuples. The objective function is somehow obtained through user preferences, for example the user can express a preference/weight for certain attributes. The main drawback of top-k queries is the fact that it is not easy for the user to specify precise preferences/weights each time and a little variation in the weights can give different results. Moreover, the users with a top-k query approach could lose some possible interesting objects in the dataset. To solve this last negative point skyline queries were introduced. Their goal is to return all objects that are not dominated by others, i.e., there exist no other tuple that is better for each attribute (where with better we can consider greater or lower values depending on the convention taken). The drawbacks of skyline queries are that they can return too many tuples if the dataset is high dimensional, that means that the number of attributes of the tuples is high, and in addition it is not possible to give any kind of preference to any attributes. Over the past few years researchers have developed new methods to solve some of these problems: Flexible Skylines is a technique that extends skylines adding constraints to attributes of objects. In particular a new concept of F-dominance is introduced, where F is a set of arbitrary families of monotone functions. Tuple t F-dominates tuple s when t is always better than or equal to s according to all the scoring functions in F (and strictly better for at least one function in F). F-skylines allow to establish a tradeoff among different attributes. They inherit from skyline queries the capability of providing an overall view of interesting results but can now focus on specific parts of the skyline, depending on the user preferences [1,2].

Another similar approach is to define a region R of preferences vectors v that are distant maximum ρ from a seed vector w that represents preferences of the user. If a record r$i$ scores at least as high as another r$j$ for every vector v in R, and strictly higher for at least one of them, we say that r$i$ ρ-dominates r$j$ . The records that are ρ-dominated by fewer than $k$ others form the ρ-skyband [3]. These notions allow to define two new operators ORD and ORU that will be described in detail in section 2.2.

A possible problem with ORD and ORU is that they require a precise seed vector w that is often not exact. A way to deal with this issue, is to expand the (explicitly input or computationally learned) weight vector into a region R and report all possible top-k sets on R partitions to the user. This task is accomplished by the uncertain top-k query (UTK) method [4].

If anyway a preference vector is not obtainable other techniques have been developed as k-representative regret minimization operator (k-regret) and k-Regret Minimizing Sets. The goal of these methods is to minimize the regret that a user could have in receiving a list of k objects [5], where the regret is defined in a precise way that is described in section 2.4

Another way of not using preference vectors is applying recursively the skyline operator on a dataset in order to obtain different skylines as done for skyline ordering where it is possible to use arbitrary size constraints while in normal skyline it is not possible [7].

One more approach extending skyline is Skyline Ranking that is a method that compares and ranks the skyline points of a dataset using a weighted directed graph, which goal is to represent the dominance relations between the skyline points [8].

All these techniques will be described in deeper detail in the following sections and then they will be compared in order to give suggestions to the reader on which approach to choose.

## 2 Description of techniques proposed
### 2.1 Flexible Skylines

| Car Id | Price($\times 10^3$) | Mileage($\times 10^3$) |
|--------|----------------------|------------------------|
| C1 | 10 | 35 |
| C2 | 18 | 25 |
| C3 | 20 | 30 |
| C4 | 20 | 15 |
| C5 | 25 | 20 |
| C6 | 35 | 10 |
| C7 | 40 | 5 |

figure 1 : table describing a dataset of cars

Flexible skylines can take into account the different importance of different attributes by means of constraints on the weights used in a scoring function. The general idea is to consider a set of scoring functions F and to introduce the concept of F-dominance: a tuple t belonging to a dataset D is said to F-dominate tuple s belonging to D when t is always better than or equal to s according to all the functions in F, where better means that the tuple t has a greater or lower score than s depending on which convention is used. From this notion of F-dominance two new operators are defined: ND and PO. ND returns all the non-dominated tuples of a dataset, considering a set of monotone functions F, while PO returns all the tuples in the dataset that are better of all the other tuples for at least one function in F. It is clear that for the same dataset and the same set of functions F, PO will return a subset of ND since if a tuple t is in PO it is not possible that t will be F-dominated by any other tuple since there exist a function in F for which it is the best tuple. Moreover if we extend F to all the monotone functions, the PO and ND operator will coincide with the skyline operator.

Let's consider an example in which we have a dataset of 8 tuples representing cars, with each tuple that is described by price and mileage. In this case that the dimension of the dataset is 2 (price,mileage) and the size of the dataset is 8 ( total number of cars). In such dataset we can evaluate each single car with a scoring function of the attributes of the cars. For example if our function is f(price,mileage)=price + mileage according to the dataset described in [figure 1](), the score of C1 will be 10 +35 = 45 and the one of C2 will be 18+25=43. We consider better in this case a lower value of the result of the score function, so in this case C2 is better than C1. Now the main concept is if instead of using a unique function, we use a set of functions as for example F = {wP Price + wM Mileage | wP ≥ wM }, in which price weighs more than mileage , to evaluate all cars that are non- F-dominated by any other car for each function in F. In this case since the function family considers more the price than the mileage is quite easy to understand that C6 and C7 are dominated by C4 since C6 and C7 have a relatively high price and ,consequently for all functions f in F, f(C6) and f(C7) will be greater than f(C4) and so C6 and C7 will be worse than C4 considering the convention adopted. In this example the only non-F-dominated tuples will be C4,C1 and C2. Out of the tuples in ND, only C1 and C4 are also part of PO; indeed C1 is the best option when wP = 0.9 and wM = 0.1, while C4 wins when weights are more comparable, e.g., when wP = 0.6 and wM = 0.4. At this point the reader may wonder what kind of families of functions we could use to compute possibly efficiently these operators. Well in [2] weighted $L_p$ norms have been proposed, defined as $L_p^W(t) = (\sum_{i=1}^{d} t[A_i]^p w_i)^{1/p}$ where W = (w1, . . . , wd) is a normalized weight vector, t is a generic tuple , $t[A_i]$ is the value of the i-th attribute of t, and p belongs to N. Later in [1] monotonically transformed, linear-in-the-weights (MLW) functions have been proposed, i.e., functions with the following form : $f^W(t)=h(\sum_{i=1}^{d} w_i g_i(t[A_i]))$ where W = (w1,...,wd ) ∈ **W**(C) where **W** is the set of all normalized weight vectors , C is a set of linear constraints, and all $g_i$ 's and h are continuous, monotone transforms such that the $g_i$ 's and h are all either (i) non-decreasing, or (ii) nonincreasing. The class of MLW functions extends weighted $L_p$ norms and covers the majority of practically relevant cases when considering monotone scoring functions.

## 2.2 ORU and ORD operators

ORU and ORD operators were introduced in [3] with the main goal of obtaining new techniques that could be output-size-specified (OSS) and at the same time to be flexible and hence not returning points considering only a single preference vector( the weights of each attribute) as happens in top-k queries.

These operators use preference vectors that multiplied by a tuple define the utility score of that tuple for that preference vector. For example considering [figure 1]() a possible tuple is C1 =(10,35) and a possible preference vector is (0.7,0.3). In this case the total score of C1 will be 10*0.7 + 0.3*35=17.5.

The main idea is to define a region of preference vectors that is arbitrary big starting from a seed vector w that represents the preferences of the user. The preference region is defined as the region in which all preference vectors v are within distance ρ (the so-called inflection radius) from the seed w (|v − w| ≤ ρ) . In practice since this region is a d-dimensional simplex, we can consider only vertices of this region as valid preferences vectors . Then if a record r scores at least as high as another s for every such vector v, and strictly higher for at least one of them, we say that r ρ-dominates s.

The records that are ρ-dominated by fewer than k others form the ρ-skyband. This general notion includes the ρ-skyline as a special case for $k = 1$. With these concepts we can describe now ORD operator: it requires in input the seed vector w , output size m and an integer k, and then it reports the records that are ρ-dominated by fewer than $k$ others, for the minimum ρ that produces exactly $m$ records in the output. The other operator ORU still requires in input the seed vector w ,the output size m and an integer k, but it reports the records that belong to the top-$k$ result for at least one preference vector within distance ρ from w, for the minimum ρ that produces exactly $m$ records in the output. For example considering the dataset in [figure 1]() , we evaluate ORD and ORU operator for m=3 ,w=(2,1) and k =2 . We start for example with for an initial $\rho = 1$ and we consider the preference vector's region R given by $|v - (0.6, 0.4)| < 1$. Starting from ORD operator we have to evaluate all the tuples with all possible vertices of R that are (3,1),(1,1),(2,2), ( 2,0)

| Car Id | <Car (3,1)> | <Car,(1,1)> | <Car ,(2,2)> | <Car , (2,0)> |
|---|---|---|---|---|
| C1 | 65 | 45 | 90 | 20 |
| C2 | 79 | 43 | 86 | 36 |
| C3 | 90 | 50 | 100 | 40 |
| C4 | 75 | 35 | 70 | 40 |
| C5 | 95 | 45 | 90 | 50 |
| C6 | 115 | 45 | 90 | 70 |
| C7 | 125 | 45 | 90 | 80 |

figure 2 : all points of figure 1 evaluated through vertices preferences vector of region R

From results of table in figure 2 we can see that C1 is dominated by (C3,C5,C6,C7) , C2 is dominated by (C3,C5,C6,C7), C3 is not dominated by anyone , C4 is dominated by (C3,C5,C6,C7), C5 is dominated by (C6,C7) , C6 is dominated by (C7), C7 is dominated by anyone. At this point the ORD operator for k=2 and m=3 will return C3,C6,C7 that are tuples dominated by zero or one tuple but it could also return C5 that is dominated by two tuples. In general it is possible to increase or decrease the inflection radius in order to respectively increase or decrease the number of results. Note that a larger $\rho$ implies a larger $\rho$-skyband. In the extreme settings, $\rho = 0$ renders the $\rho$-skyband equivalent to a traditional top-$k$ query at w, while $\rho = \infty$ makes it equivalent to the standard $k$-skyband.

Considering the results of figure 2 we can evaluate ORU for $\rho = 1$ , m=3 , k=2. We have to search all tuples that rank in top-2 position for at least one preference vector : for vector (3,1) results are (C7,C6) , for (1,1) we have (C3 and C6 tied with C7,C1,C5) , for (2,2) we have (C3 and C6 tied with C7,C5,C1) and for (2,0) we have (C7,C8). In this case basically we have found all the dataset except C4 and C2. We can select 3 tuples from the list of tuples obtained to be returned. What both ORU and ORD do in practice is to calculate the minimum $\rho$ for which they return exactly m tuples.

## 2.3 Uncertain Top-k Queries (UTK)

Uncertain top-k queries (UTK) operators ( UTK1 and UTK2, two different variants) were introduced in [4] with the aim of obtaining operators able to manage uncertainty in the preference vector and at the same time to provide the user with a variable number of results based on similar preference profiles. Previously research on top-k processing has only considered uncertainty in the data ( [9,10]), but not in the weight vector. The data uncertainty typically assumed is at the record or at the attribute level [11]. In the former case, each record may exist in the dataset with a certain probability. In the latter, the score of every record is a random variable that follows a given distribution. In both cases, there is no uncertainty in the weight vector, and the top-k answer is probabilistic. Also, the records' scores are independent. In contrast, there is nothing probabilistic in UTK (e.g., the output is exact depending only on the input).

Both UTK operators require in input a dataset D, a positive number k , and a region R. The main idea is to consider the region R as the preference domain of the user, obtained asking the user a generic preference vector that is then expanded in a region or for example asking the user to input a range of weights for each attribute. For example if a user is interested in hotels that are ranked by Service, Cleanliness and Location, she could input a generic preference vector (that will be expanded somehow by the application in a region, infact UTK requires the region directly not a weight vector to expand) or directly a range for attributes' weights for example [0.05, 0.45] for Service and [0.05, 0.25] for Cleanliness. The user doesn't need to specify weights also for Location since UTK consider that the sum of all components of each preference vector is 1. Consequently it is possible to reduce the domain of preference vectors to a (d − 1)-dimensional space where d is the total number of attributes.

That said UTK1 reports the set of exactly those records that may rank among the top-k for a generic weight vector that lies inside R. "Exactly" here means that the reported set is minimal, i.e., for every record p in it, there is at least one weight vector in R for which p belongs to the top-k set. The second version, UTK2, reports the specific top-k records for any possible positioning of preferences vectors in the region. While there are infinite possible vectors in R, the output is a partitioning of R, where each partition is associated with the exact top-k set when w lies anywhere inside that partition.

| Hotel | Service | Cleanliness | Location | total score for preference vector (0.33,0.33,0.33) |
|---|---|---|---|---|
| p1 | 8.3 | 9.1 | 7.2 | 8.2 |
| p2 | 2.4 | 9.6 | 8.6 | 6.86 |
| p3 | 5.4 | 1.6 | 4.1 | 3.7 |
| p4 | 2.6 | 6.9 | 9.4 | 6.3 |
| p5 | 7.3 | 3.1 | 2.4 | 4,26 |
| p6 | 7.9 | 6.4 | 6.6 | 6.76 |
| p7 | 8.6 | 7.1 | 4.3 | 6.66 |

figure 3 : a dataset of hotels considered by service, cleanliness and location

Considered the dataset in figure 3, for k= 2 and the region R expressed by the axis-parallel rectangle e[0.05, 0.45] × [0.05, 0.25] UTK1 returns {p1,p2,p4,p6} that are the vectors that are in top-2 for a least one preference vector in R. Intuitively, as we can see from the figure, for a preference vector that evaluate attributes of records with the same weight, p1,p2 and p6 have high scores and in fact we can find them also in results of UTK1, but we can find also p4 that has a lower value for example than p7 for preference vector(0.33,0.33,0.33). This is due to the fact that R contains preference vectors with the location weight that is greater than the service one or the cleanliness one because it was defined for low ranges of values for Service and Cleanliness weights. Infact p7 in R never ranks at least second due to his low score on Location while p4 has a really good score in Location.

Considering UTK2 for this example, it divides R in partitions identifying them by the records that are top-2 for each preference vector in the partition. In this case UTK2 identify 4 region (1 with weight on Service, identified by p2,p4 , one with high weight on Service identified by p1,p6) and two regions for intermediate weight on Service and higher /lower weight for Cleanliness, identified respectively by p1,p2 and p1,p4). Depending on how many results the user wants UTK2 could return a different number of this tuples characterized by similar preference profiles.

## 2.4 Regret minimization operators

the k-representative regret minimization operator (k-regret) was introduced in [5] to attempt for an operator that has features from both top-k and skyline. That is, an operator that outputs a small set of k tuples without asking utility functions. The main idea is based on happiness of a user seeing k result from her query. Given a list of k tuples, we say that a user is x% happy with the list if the utility she obtains from the best tuple in this list is at least x% of the utility she obtains from the best tuple in the whole database. The goal of k-regret operator is to maximize this happiness taking in consideration that utility is measured through the class of linear functions. More formally we define the gain of a user as the maximum score she gets from a subset S of points belonging to a dataset D and considering a particular function. For example taking in consideration the dataset D of figure 4 , we can select a subset S = {p1,p2} and a function $f_{(0.2,0.8)}$ and evaluate the gain(S, f) = $max_{p \in S}$ f(p) = 117.4 (achieved by p1)

That said we can define the regret as $r_D$(S, f) = gain(D, f)−gain(S, f) and the regret ratio $rr_D$(S, f) = $\frac{r_D(S,f)}{gain(D,f)}$ . In our example we have $r_D$(S, f) = 165.4−117.4 = 48 and regret ratio $rr_D$(S, f) = 48/165.4 = 0.29

| Car | MPG | HP | $f_{(0.2,0.8)}$ | $f_{(0.4,0.6)}$ | $f_{(0.6,0.4)}$ | $f_{(0.8,0.2)}$ |
|---|---|---|---|---|---|---|
| p1 | 51 | 134 | 117.4 | 100.8 | 84.2 | 67.6 |
| p2 | 40 | 110 | 96 | 82 | 68 | 54 |
| p3 | 41 | 191 | 161 | 131 | 101 | 71 |
| p4 | 35 | 198 | 165.4 | 132.8 | 100.2 | 67.6 |
| p5 | 30 | 140 | 118 | 96 | 74 | 52 |

figure 4: a dataset of cars considered by high miles per gallon and high horse power, with their scores for a class F = { $f_{(0.2,0.8)}$, $f_{(0.4,0.6)}$, $f_{(0.6,0.4)}$, $f_{(0.8,0.2)}$ }

Let the class of utility functions being considered be F. We now define the worst possible regret for any user with a utility function in F as $rr_D(S, F) = sup_{f \in F} rr_D(S, f)$. Continuing the example of figure 4 with S={p1,p2} and F={$f_{(0.2,0.8)}$, $f_{(0.4,0.6)}$, $f_{(0.6,0.4)}$, $f_{(0.8,0.2)}$}, the maximum regret ratio is 0.29. To see this, we use values in the table to get the regret ratios $rr_D$ (S, f(0.2,0.8)) = 0.29, $rr_D$ (S, f(0.4,0.6)) = 0.24, $rr_D$ (S, f(0.6,0.4)) = 0.17, and $rr_D$ (S, f(0.8,0.2)) = 0.05. In general the goal of the k-regret operator, given the number of tuples k to return, is to return a set of k tuples such that the maximum regret ratio is minimized. Considering the example considered so far the result for k=2 would be {p3,p4} with maximum regret = $\frac{f_{(0.2,0.8)}(p4) - f_{(0.8,0.2)}(p4)}{f_{(0.2,0.8)}(p4)}$ = (165.4 – 67.6) /165.4 = 0.59 . If we try to substitute any of p3,p4 with another tuple, the maximum regret will increase since the other tuples have at least one function in F for which they score worse and this lead to a greater maximum regret.

A possible drawback of k-regret operator is that it always takes in consideration the best tuple of the dataset to evaluate the maximum regret ratio. This could be a problem when the best tuple of the dataset doesn't really suit the user's preferences especially when the bust tuple is an outlier with other tuples that has values of attributes quite different.

To overcome this drawback, k-regret minimizing sets operator was introduced in [6] as a generalization of the k-regret operator. It measures how far from a k'th "best" tuple of the dataset D is the "best" tuple in a subset S. More formally it defines the kgain as the score of the k-ranked point in a subset R for a preference vector. Then it defines the k-regret ratio for a particular preference vector w as k-regratio(S, w) = $\frac{max(0, kgain(D,w) - 1gain(S,w))}{kgain(D,w)}$ .To notice the fact that the value must fall in range [0,1]. For example considering the dataset in figure 4 for S={p1,p2,p5} and $f_{(0.2,0.8)}$ that is equivalent to w= (0.2,0.8) , for k = 2 we have the 2-regratio({p1,p2,p5},(0.8,0.2) ) = max(0, 161-118)/161 = 0.267 , where 161 is the score of the second best tuple in the dataset which is p3 while 118 is the score of the best tuple in S which is p5.

That said we can define the maximum k-regret ratio for a subset S⊆D and for a class of preference vectors L, as the greater k-regret ratio evaluated for R and for all possible preference vector in S. Considering our example we found out previously that $2 - regratio(\{p1, p2, p5\}, (0.8, 0.2)) = 0.267$. Similarly $2 - regratio(\{p1, p2, p5\}, (0.4, 0.6) = max (0, 131 - 118)/131 = 0.099$,
$2 - regratio(\{p1, p2, p5\}, (0.6, 0.4) = \frac{max(0, 100.2 - 84.2)}{100.2} = 0.159$, $2\ regratio(\{p1, p2, p5\}, (0.8, 0.2) = max(0, 67.6 - 67.6)/67.6 = 0$ and consequently 0.267 is the maximum regret ratio for this example.

At this point we focus on the primary objective of this technique, to produce a fixed-size k-regret minimizing set. Given an integer r and a dataset D, discover a subset S ⊆ D of size r that achieves the minimum possible maximum k-regret ratio given a class of weight vectors L. Considering our example of figure 4, if we have r=2 and k=2 in input and considering only tuples p1,p2,p3.p4 in the dataset D to avoid many calculations , we have to discover which set among {p1,p2},{p1,p3},{p3,p2}, {p1,p4},{p2,p4},{p3,p4} has the minimum maximum regret. Doing some calculus as we have done before , we find that that the maximum 2-regret ratio for {p1,p2} is 0.27 , for {p1,p3} is 0 , for {p3,p2} is 0, for {p1,p4}is 0, for {p2,p4} is 0 , {for p3,p4} is 0. So basically this mean that tuple p3 and p4 guarantee the same utility than the second best tuple in the dataset , and therefore a set that contains one of them has 0 as maximum regret-ratio and could be a result for the 2-regret minimizing sets operator.

## 2.5 Skyline Ordering

The traditional skyline operator doesn't support size constraints on queries. To overcome this issue a technique called pointwise ranking was introduced with the goal of resolving the query size constraint by ranking all points with a dominance-based scoring function. An example of such function is to count for a point p, the number of points it dominates [12]. Another alternative is set-wide maximization that instead of considering each single point separately, considers a set of points collectively to maximize a target value. A possible approach with this method is to retrieve a subset S of points from a dataset D such that the number of points in P \ S that are dominated by some point in S is maximized [13]. Both this approaches have limitations: for pointwise optimization as we count dominating points only, non skyline points may be preferred over skyline points. Moreover it can incur in too many ties, which invalidates the ranking. A drawback of the set-wide maximization described is that it is not applicable to arbitrary size constraints on skyline queries because it is unable to handle the case where more points than the skyline cardinality are expected. To overcome this issues Skyline Order operator was defined in [7] . The skyline order of a dataset D is a sequence S =<$S_1, S_2, ...., S_n$> where  $S_1$ is the skyline of  P and ∀i , 1<i<=n $S_i$, is the skyline of D \ $\cup_{j=1}^{i-1} S_j$. Referring to each skyline in the skyline order as sub-skyline, we have that, by construction, the union of points of all skyline is D and there are no points in common between two different sub-skylines. Moreover given the skyline order <$S_1, S_2, ...., S_n$> of a d-dimensional point set D, the skyline order index of a point p ∈ P is i if p ∈ $S_i$. Said that, all the sub-skylines with an order smaller than another sub-skyline S contains at least one point that dominates all points in S.

This lead to a sort of ranking in the skyline order: depending on the size of the result requested, it is possible to return points from more than one sub-skyline starting from the ones with smaller order.

For example we want to calculate the skyline order of the dataset D of figure 3. $S_1$  will be the traditional skyline of the dataset {p1,p2,p4,p7}. At this point we consider D \ $S_1$  = {p3,p5,p6} and we use it  to calculate the skyline again to compute $S_2$ that is  {p6}. Now we consider D \ ( $S_1$ ∪ $S_2$) ={p3,p5} for computing $S_3$={p3,p5}. If the requested input size is 5 the skyline order operator will return all the points in $S_1$ plus one more from $S_2$.

## 2.6 Skyline Ranking

A typical problem of skyline operator is that it returns too many tuples as the dimension of the dataset is big. A possible way to solve this problem is doing Skyline ranking, a technique that tries to give a score at the different tuples of the skyline of a dataset.

In particular we focus on Skyrank, a method introduced in [8].In order to understand it  we have to introduce some concepts : formally, given a data space S defined by a set of d dimensions {$d_1, ..., d_n$} and a dataset D on S, a point p ∈ D  is represented as p= {$p_1, ..., p_n$} where $p_i$, is a value on dimension $d_i$. Each non-empty subset U of S (U ⊆ S) is referred to as a subspace of D. The data space D is also referred to as full space of the dataset S [14]. Given the definition of subspace , we can extend the concept of dominance to subspaces simply considering when checking the dominance condition not all attributes but only the ones included in the sub-space. For example considering figure 3 , we have that the space S on which the dataset D is defined is {Service,Cleanliness,Location}. A possible subset of S is $S_1$ = {Service,Cleanliness} or   $S_2$ ={Service}.If we consider greater values as  better , we have that p4 is dominated by p1 for the subset $S_1$   but it is not dominated according to S, since p4 has a better location value. Moreover we define as parent $S_p$ of a subset $S_c$ a subset which contains the same dimensions of $S_c$ plus one. $S_c$ is called child of $S_p$.For example S is a parent of $S_1$ and $S_1$ is the child of S , but $S_2$ is not the child of S and S is not a parent of $S_2$.

Skyrank is characterized by two steps: the first one builds a graph in which all nodes are the skyline points of the dataset D of state space S and the edges represent dominance between tuples according to different sub-spaces. The second step is the ranking of the tuples included in the graph.

For what concerns the first step the idea is to consider all possible non-empty subspaces , the so called skycube [15]. For all subspaces $S_B$ in the skycube the algorithm examines the dominance relation between any point in $S_B$ and the skyline points of any child subspace Y of $S_B$. For each child subspace Y where p is a skyline point for Y and S, no edge is added to the skyline graph. For each child subspace Y where p is not a skyline point for Y and S, we retrieve all skyline points q that dominate p in subspace Y and add an edge $e_{pq}$ to the skyline graph. If the edge $e_{pq}$ already exists, we just increase the number of occurrences of the edge $e_{pq}$ by one. After all subspaces of the skycube are examined, we normalize the edges' occurrences and give to each edge $e_{pq}$ a weight defined as the total number of occurrences of. $e_{pq}$ divided by the number of all edges.

For example , we can build such a graph for the dataset of figure 3. The nodes of the graph will be the skyline points of the dataset according to the state space ={Service,Cleanliness,Location} that are p1,p2,p4,p7. We start selecting {Service,Cleanliness} as first subspace in which we identify p1,p2,p7 as skyline points. Then we consider the children of {Service,Cleanliness} that are {Service} and {Cleanliness}. For {Service } the only skyline point is p7 while for {Cleanliness} is p2. As p7 dominates p1 and p2 in {Service} and p2 dominates p1,p7 in {Cleanliness}. Therefore we add $e_{p1p7}$ $e_{p2p7}$ $e_{p1p2}$ $e_{p7p2}$ as edges with a initial number of occurrences of 1. And we proceed in this way until we have analyzed all sub-spaces. At this end the result is shown in figure 5 , where the picture on the left represents the graph with the number of occurrences as weight for each edge and in the right there is the graph with the updated weight for each edge obtained dividing the number of occurrences of the edge with the total numbers of edges.

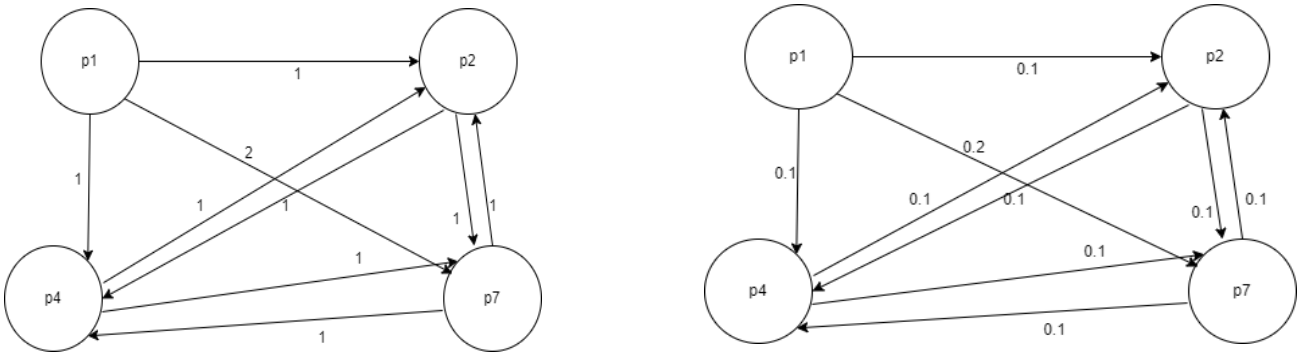

Figure 5 : skyrank graph of the dataset of figure 3

Once obtained the skyline graph, we proceed to rank all nodes of the graph. This job can be done through a link-based ranking algorithm, that is based on the concept that each vertex ,which points to another vertex through an edge ,transfers some of its authority to the linked vertex. . By using this kind of algorithm, skyline points that dominate many other skyline points which in turn dominate other points in some subspaces, are highly ranked.

An example of such algorithm is PageRank[16], one of the most popular link analysis algorithms and is broadly used by web search engines, in order to rank retrieved web pages. The basic idea of PageRank is that if page p has a link to page p', then some authority of p is implicitly transferred to page p'. In a similar way Skyrank considers a skyline point more interesting, if it dominates many other skyline points in many subspaces, and these points in turn also dominate many points in other subspaces. Skyrank defines the score of a skyline point p based on its interestingness as I(p) = (1- α) $\sum_{\forall p \in \text{domd}(q)} \frac{1}{|dom(p)|} I(q) + \alpha \frac{1}{|SKY|}$ where α∈[0, 1.0] where we denote the set of skyline points that are dominated by p in some subspace as domd(p) and the set of skyline points that dominate a skyline point p in some subspace as dom(p). The first part of the expression ($\sum_{\forall p \in \text{domd}(q)} \frac{1}{|dom(p)|} I(q)$) expresses the fact that interestingness of the dominated points is transferred to the dominating points. While the second part ($\frac{1}{|SKY|}$) indicates that all skyline points share an equal (small) amount of interestingness, since they are qualified as skyline points. The variable α is used to balance between the two parts of the score.

# 3 Comparison of the proposed techniques

All the new technique proposed try to improve top k-queries and skylines queries overcoming to their main drawbacks such as simplicity of formulation and overall view of interesting results for top-k query and control of result cardinality and trade-off among attributes for skyline queries.

The flexible skyline method proposed, improves traditional skylines returning a more specific region of the skyline , hence reducing the size of the output that was a big limitation for traditional skyline in high dimensional dataset. Anyway a drawback of F-skyline operator is that however it isn't ouput size specified since it doesn't give guarantees on the size of the output. Moreover it works with class of functions that are not easy to be chosen by the user. The user may insert some weights and then the application can create a class of function to input to a F-skyline calculator, but with this process some precision in the results could be lost.

A similar approach to F-skylines are the ORU-ORD operators that aim as well as F-skylines to be very precise to return a particular region of tuples. Their main drawback is that they require a seed vector from the user that represents her preferences. Such seed vector in general could be imprecise and non-representative of the real preferences of the user. This can lead in results imprecision. A positive aspect of ORU/ORD in relation to F-skylines is that they are output size specified.

UTK operators instead, sacrifice some precision in the result set, but they allow some uncertainty in the weight vector. UTK1 doesn't give any guarantee on the size of the result while UTK2 is more elastic since it divides the result in partitions that can be returned as different results for similar preference profiles.

Another possible approach is using regret-minimizing operators. They are output-size specified and precise once defined the parameter k, which allows to measures how far from a k'th "best" tuple of the dataset D is the "best" tuple in a subset of D. This leads to different results for different parameters k, hence it is possible to receive a more complete view of the dataset running these operators for different parameters, increasing however the complexity of the application as main drawback.

Differently from  F-skyline , ORU/ORD and UTK operators , regret-minimizing operators don't consider a different weight among attributes.

In this case similarly to regret-minimizing operators, skyline ordering and skyline ranking don't manage the possibility of different weights among attributes. They are an extension of traditional skyline that achieve a more flexible approach  for the size of the result.

The difference between the two is how the tuples are ranked, in skyline ordering the tuples contained in a sub-skyline with a smaller order are considered better, while skyline ranking gives score to each skyline point according to a certain function, evaluated on a graph in the case of Skyrank.

 In figure 6 is shown an overview of different aspects of these techniques.

|  | F-Skylines | ORD/ORU | UTK | Regret-minimazing | Skyline order | Skyline Ranking |
|---|---|---|---|---|---|---|
| Overall view of interesting results | No | No | Yes | Yes | Yes | Yes |
| Precision in results according to some functions/preferences | Yes | Yes | No | No | No | No |
| Output size specified | No | Yes | Yes for UTK2 | Yes | Yes | Yes |
| Possibility of different weights on attributes | Yes | Yes | Yes | No | No | No |
| Importance of some preferences inserted by the user | Yes | Yes | No | No | No | No |

Figure 6 : overview of the comparison of the techniques proposed

# References

[1] Paolo Ciaccia, Davide Martinenghi: Flexible Skylines: Dominance for Arbitrary Sets of Monotone Functions. ACM Trans. Database Syst. 45(4): 18:1-18:45 (2020)

[2] Paolo Ciaccia, Davide Martinenghi: Reconciling Skyline and Ranking Queries. Proc. VLDB Endow. 10(11): 1454-1465 (2017)

[3] Kyriakos Mouratidis, Keming Li, Bo Tang: Marrying Top-k with Skyline Queries: Relaxing the Preference Input while Producing Output of Controllable Size. SIGMOD Conference 2021: 1317-1330

[4] Kyriakos Mouratidis, Bo Tang: Exact Processing of Uncertain Top-k Queries in Multi-criteria Settings. Proc. VLDB Endow. 11(8): 866-879 (2018)

[5] Danupon Nanongkai, Atish Das Sarma, Ashwin Lall, Richard J. Lipton, Jun (Jim) Xu :Regret-Minimizing Representative Databases. Proc. VLDB Endow. 3(1): 1114-1124 (2010)

[6] Sean Chester , Alex Thomo , S. Venkatesh, Sue Whitesides: Computing k-Regret Minimizing Sets. Proc. VLDB Endow. 7(5): 389-400 (2014)

[7] Hua Lu, Christian S. Jensen, Zhenjie Zhang:Flexible and Efficient Resolution of Skyline Query Size Constraints. IEEE Trans. Knowl. Data Eng. 23(7): 991-1005 (2011)

[8] Akrivi Vlachou, Michalis Vazirgiannis: Ranking the sky: Discovering the importance of skyline points through subspace dominance relationships. Data Knowl. Eng. 69(9): 943-964 (2010)

[9] Y. Tao, D. Papadias, X. Lian, and X. Xiao. Multidimensional reverse k NN search. VLDB J., 16(3):293–316, 2007.

[10] L. Antova, T. Jansen, C. Koch, and D. Olteanu. Fast and simple relational processing of uncertain data. In ICDE, pages 983–992, 2008

[11] [G. Cormode, F. Li, and K. Yi. Semantics of ranking queries for probabilistic data and expected ranks. In ICDE, pages 305–316, 2009

[12] D. Papadias, Y. Tao, G. Fu, and B. Seeger, "An Optimal and Progressive Algorithm for Skyline Queries," Proc. SIGMOD, pp. 467-478, 2003.

[13] X. Lin, Y. Yuan, Q. Zhang, and Y. Zhang, "Selecting Stars: The k Most Representative Skyline Operator," Proc. Int'l Conf. Data Eng. (ICDE), pp. 86-95, 2007.

[14] J. Pei, W. Jin, M. Ester, Y. Tao, Catching the best views of skyline: a semantic approach based on decisive subspaces, Proceedings of International Conference on Very Large Data Bases (VLDB), 2005
[15] Y. Yuan, X. Lin, Q. Liu, W. Wang, J.X. Yu, Q. Zhang, Efficient computation of the skyline cube, Proceedings of International Conference on Very Large Data Bases (VLDB), 2005
[16] S. Brin, L. Page, The anatomy of a large-scale hypertextual Web search engine, Computer Networks 30 (1–7) (1998)